\documentclass{sf2a-conf2010}
\usepackage{graphicx}
\usepackage{hyperref}
\usepackage[]{natbib}

\providecommand\apjs{ApJSupp}                 
\providecommand\aap{A\&A}            

\providecommand{\eprint}[1]{\href{http://arxiv.org/abs/#1}{{\tt [arXiv:#1]}}}
\providecommand{\url}[1]{\href{#1}{#1}}

\newcommand\ddeg{\ifmmode^\circ\else$^\circ$\fi}    

\begin{document}
\TitreGlobal{SF2A 2010}
%
\title{Galactic Plane image sharpness as a check on cosmic microwave background mapmaking}
\author{Boudewijn F. Roukema}\address{Toru\'n Centre for Astronomy, Nicolaus Copernicus University, ul. Gagarina 11, 87-100 Toru\'n, Poland}
%
\runningtitle{Galactic Plane CMB mapmaking check}
%
\setcounter{page}{237}
\index{Roukema, B.F.}

\maketitle
\begin{abstract}
%
The largest uncollapsed inhomogeneity in the observable Universe is
statistically represented in the quadrupole signal of the cosmic
microwave background (CMB) sky maps as observed by the Wilkinson
Microwave Anisotropy Probe (WMAP).  The constant temporal offset of
$-25.6$~ms between the timestamps of the spacecraft attitude and
observational data records in the time-ordered data (TOD) of the WMAP
observations was suspected to imply that previously derived all-sky
CMB maps are erroneous, and that the quadrupole is in large part an
artefact. The optimal focussing of bright objects in the Galactic
Plane plays a key role in showing that no error occurred at the step
of mapmaking from the calibrated TOD. Instead, the error had an effect
when the uncalibrated TOD were calibrated. Estimates of the
high-latitude quadrupole based on the wrongly calibrated WMAP maps are
overestimated by about 15--60\%.
\end{abstract}
%
  

Although the primary aim of cosmic microwave background (CMB) all-sky
observation missions is cosmological, the Galaxy constitutes a major
component of the resulting data set. \nocite{LL09lowquad}{Liu} \& {Li} (\protect\hyperlink{hypertarget:LL09lowquad}{2010}) reconstituted
all-sky maps from Wilkinson Microwave Anisotropy Probe (WMAP,
\nocite{WMAPbasic}{Bennett} {et~al.} \protect\hyperlink{hypertarget:WMAPbasic}{2003})
time-ordered data (TOD) and suggested that the quadrupole present in
the official versions of the maps is mostly an artefact, since their own
maps had a weaker quadrupole. They later traced this to a timing
offset of $-25.6$~ms between the timestamps of the spacecraft attitude
and observational data records in the calibrated TOD files
\nocite{LL10toffset}({Liu} {et~al.} \protect\hyperlink{hypertarget:LL10toffset}{2010}).  Since the offset is also present in the uncalibrated
TOD files, it could have affected either (i) the calibration step or
(ii) the mapmaking step.

\begin{figure}[ht!]
 \centering
 \includegraphics[width=0.8\textwidth]{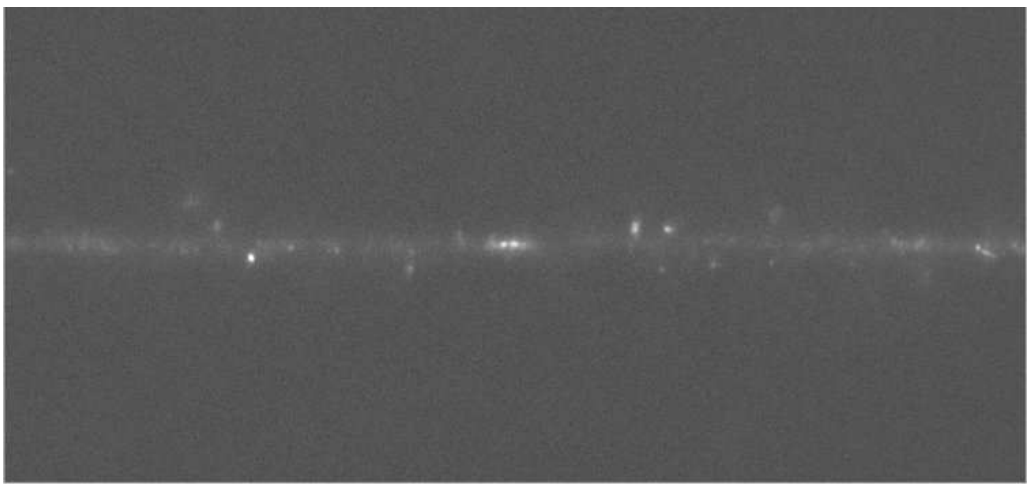}
  \caption{Correctly focussed \protect\nocite{Rouk10Toffa}($\delta t=0.5$,  {Roukema} \protect\hyperlink{hypertarget:Rouk10Toffa}{2010a}) but wrongly
    calibrated \protect\nocite{Rouk10Toffb}({Roukema} \protect\hyperlink{hypertarget:Rouk10Toffb}{2010b}) WMAP W~band (94~GHz) image of the $53.0\ddeg
    \times 24.7\ddeg$ region centred at the Galactic Centre (North up,
    East left), after monopole and dipole subtraction, on a grey scale
    ranging from black (-20~mK) to white (+40~mK). To zoom in, see Fig.~4, 
    \protect\nocite{Rouk10Toffa}{Roukema} (\protect\hyperlink{hypertarget:Rouk10Toffa}{2010a}).}
  \label{roukema:fig1}
\end{figure}

\begin{figure}[ht!]
 \centering
 \includegraphics[width=0.8\textwidth]{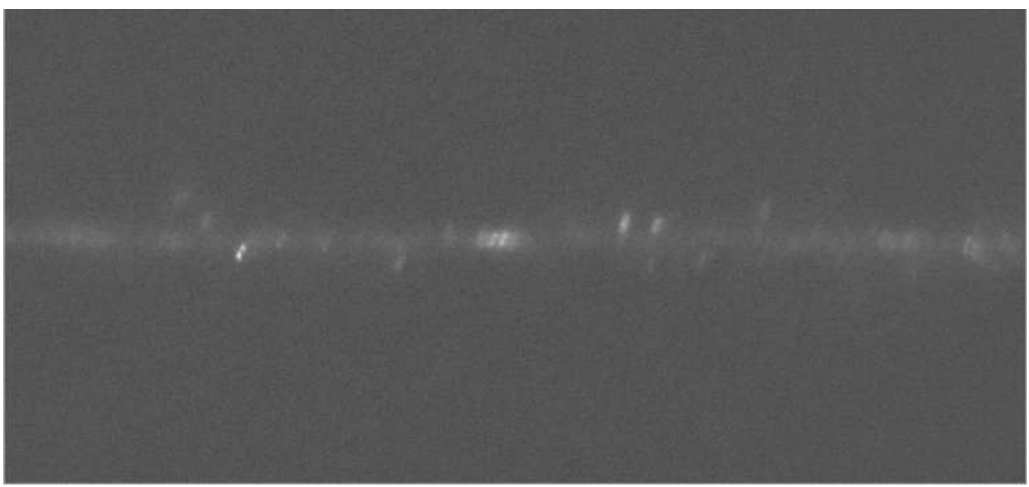}
  \caption{Wrongly focussed, wrongly calibrated WMAP W band image, as
    for Fig.~\protect\ref{roukema:fig1}, with a timing offset $\delta
    t=-5$, i.e. exaggerated by a factor of ten beyond that which
    generated the calibration error. To zoom in, see Fig.~2, 
    \protect\nocite{Rouk10Toffa}{Roukema} (\protect\hyperlink{hypertarget:Rouk10Toffa}{2010a}).}
  \label{roukema:fig2}
\end{figure}

The WMAP 3-year calibrated TOD were compiled into maps using \nocite{LL10toffset}{Liu} {et~al.} (\protect\hyperlink{hypertarget:LL10toffset}{2010})'s 
publicly available data analysis pipeline\footnote{\url{http://cosmocoffee.info/viewtopic.php?p=4525}, 
\href{http://dpc.aire.org.cn/data/wmap/09072731/release_v1/source_code/v1/}{{\tt http://dpc.aire.org.cn/data/wmap/09072731/release\_v1/source\_code/v1/}}}, 
and patched for using the GNU Data Language (GDL) and for two different
timing error tests. In both cases, the timing offset, written as a multiple 
$\delta t$ of an exposure time in a given waveband, where $\delta t =0.5$ 
corresponds to the timing offset used by the WMAP collaboration, was varied 
in order to detect its effect on a relevant statistic of the maps.
Testing an error at step (i) was done by creating
low-resolution maps and finding the maps with the least variance per pixel \nocite{Rouk10Toffb}({Roukema} \protect\hyperlink{hypertarget:Rouk10Toffb}{2010b}).\footnote{\href{http://cosmo.torun.pl/GPLdownload/LLmapmaking_GDLpatches/LLmapmaking_GDLpatches_0.0.4.tbz}{{\tt http://cosmo.torun.pl/GPLdownload/LLmapmaking\_GDLpatches/LLmapmaking\_GDLpatches\_0.0.4.tbz}}}
Testing an error at step (ii) was done by calculating
high-resolution maps that included sub-cosmological objects, and finding the best
focussed maps \nocite{Rouk10Toffa}({Roukema} \protect\hyperlink{hypertarget:Rouk10Toffa}{2010a}).\footnote{\href{http://cosmo.torun.pl/GPLdownload/LLmapmaking_GDLpatches/LLmapmaking_GDLpatches_0.0.3.tbz}{{\tt http://cosmo.torun.pl/GPLdownload/LLmapmaking\_GDLpatches/LLmapmaking\_GDLpatches\_0.0.3.tbz}}}  
The results, summarised in Table~\ref{roukema:tab1}, showed to very high
significance that the error affected the calibration step, but did not affect
the mapmaking step directly. However, maps made from the wrongly calibrated 
data necessarily include the calibration error.
For example, estimates of the
high-latitude quadrupole based on the wrongly calibrated WMAP maps are
overestimated by about 15--60\% \nocite{Rouk10Toffb}({Roukema} \protect\hyperlink{hypertarget:Rouk10Toffb}{2010b}). Figures~\ref{roukema:fig1}
and \ref{roukema:fig2} illustrate the sharpest focus test.


\begin{table}[b]
  \centering
  \caption{ Comparison of sharpest focus 
    and minimum variance methods of testing for a timing offset error.
    \label{t-cf_two_methods}}
  \begin{tabular}{c c c c cc } \hline
    \rule[-1.5ex]{0ex}{4.5ex}
    short name & minimum variance & sharpest focus \\
    reference & \protect\nocite{Rouk10Toffb}{Roukema} (\protect\hyperlink{hypertarget:Rouk10Toffb}{2010b}) & \protect\nocite{Rouk10Toffa}{Roukema} (\protect\hyperlink{hypertarget:Rouk10Toffa}{2010a}) \\
    step to understand & uncal.\ TOD $\rightarrow$ cal.\ TOD
    & cal.\ TOD $\rightarrow$ map \\
    step analysed & cal.\ TOD $\rightarrow$ map 
    & cal.\ TOD $\rightarrow$ map \\
    planets \& Gal.\ Plane & excluded & included \\
    $N_{\mathrm{side}}$ & 8 & 2048 \\
    statistic  & variance per pixel & brightness of 503-rd brightest pixel \\
    max/min  & min & max \\
    rejected hypothesis & $\delta t = 0.5$ rejected at 8.5$\sigma$ &
     $\delta t = 0$ rejected at 4.6$\sigma$ \\
    accepted hypothesis & $(\delta t-0.5) \times 52.1$~ms = $-25.6$~ms & $\delta t =0.5$ \\
    conclusion & calibration step wrong & mapmaking step right \\
    \hline
  \end{tabular}
  \label{roukema:tab1}
\end{table}

%
%
%
%
%


%

\begin{thebibliography}{}

\bibitem[{Bennett}, {Halpern}, {Hinshaw}, {Jarosik},  {Kogut}, {Limon}, {Meyer}, {Page}, {Spergel}, {Tucker}, {Wollack}, {Wright},  {Barnes}, {Greason}, {Hill}, {Komatsu}, {Nolta}, {Odegard}, {Peiris},  {Verde}, \& {Weiland} (2003)]{WMAPbasic}
\hypertarget{hypertarget:WMAPbasic}{{Bennett}, C.~L., {Halpern}, M., {Hinshaw}, G., {et al.} 2003,} \apjs, 148, 1,  \eprint{astro-ph/0302207}

\bibitem[{Liu} \& {Li} (2010)]{LL09lowquad}
\hypertarget{hypertarget:LL09lowquad}{{Liu},} H., \& {Li}, T. 2010, ArXiv e-prints, \eprint{0907.2731}

\bibitem[{Liu}, {Xiong}, \& {Li} (2010)]{LL10toffset}
\hypertarget{hypertarget:LL10toffset}{{Liu}, H., {Xiong}, S., \& {Li}, T. 2010, ArXiv e-prints,} \eprint{1003.1073}

\bibitem[{Roukema} (2010a)]{Rouk10Toffa}
\hypertarget{hypertarget:Rouk10Toffa}{{Roukema}, B.~F. 2010a,} \aap, in press, \eprint{1004.4506}

\bibitem[{Roukema} (2010b)]{Rouk10Toffb}
\hypertarget{hypertarget:Rouk10Toffb}{{Roukema}, B.~F. 2010b,} \aap, submitted, \eprint{1007.5307}

\end{thebibliography}
\end{document}